\documentclass[singlecolumn]{jpsj3}
\usepackage[left=2cm,top=2cm,right=2cm, bottom=2cm]{geometry}
\usepackage{txfonts}
\usepackage{graphicx}
\usepackage{cite}
\usepackage{overcite}

\title{Coexistence of the Pseudogap and the Superconducting Gap Revealed by the \textit{c}-axis Optical Study of YBa$_{2}$(Cu$_{1-x}$Zn$_{x}$)$_{3}$O$_{7-\delta}$}

\author{\name{Ece \surname{Uykur}}\thanks{Email:uykure@tsurugi.phys.sci.osaka-u.ac.jp}, \name{Kiyohisa \surname{Tanaka}}, \name{Takahiko \surname{Masui}}, \name{Shigeki \surname{Miyasaka}}, \name{Setsuko \surname{Tajima}}}
\inst{Department of Physics, Graduate School of Science, Osaka University, Osaka, 560-0043, Japan} 

\abst{To address the issue of the pseudogap in the superconducting state of the high temperature superconducting cuprates, we studied the temperature dependent redistribution of the spectral weight in the \textit{c}-axis polarized optical spectra for underdoped YBa$_{2}$(Cu$_{1-x}$Zn$_{x}$)$_{3}$O$_{7-\delta}$ single crystals with $x=0,0.007,$ $0.012$, and $0.04$. It was found that for Zn substituted samples, the spectral weight transfer to the high energy region, which characterizes the pseudogap state, continues even below the superconducting transition temperature (\textit{T}$_{c}$), indicating the coexistence of the pseudogap and the superconducting gap. Moreover, the observation of the pseudogap in the Zn-doped non-superconducting sample gives evidence that the pseudogap is not a precursor of superconductivity. }

\kword{YBa$_{2}$Cu$_{3}$O$_{7-\delta}$, c-axis optical spectra, Zn-substitution effects, pseudogap}

\begin{document}
\maketitle

Since the discovery of high-\textit{T}$_{c}$ cuprate superconductors, the aberrant normal state properties of these materials have been under intensive debate. Especially, the pseudogap phenomenon, which strongly dominates the underdoped side of the electronic phase diagram, has long been a topic of discussion. There are several models that try to explain the pseudogap and its relation to the superconducting gap. Some of them explain the pseudogap as a precursor of superconductivity that reflects the pair fluctuations above \textit{T}$_{c\text{ }}$\cite{scenario1a, scenario1b}, while in some other theories, the pseudogap is a competing order such as antiferromagnetic order or spin/charge density wave, etc \cite{scenario2a, scenario2b, scenario2c}.  

The pseudogap state has been observed by many spectroscopic probes as discussed in Ref. 6. 
Optical spectroscopy is an especially powerful probe because it can clearly distinguish a superconducting gap from the insulating gaps. In particular, the \textit{c}-axis polarized optical spectrum is very sensitive to the electronic density of states in the antinodal region \cite{caxisisantinodal, matrix element} of the Fermi surface near ($0,\pi$) and ($\pi,0$), where the pseudogap has a strong effect, while a major pseudogap effect on the in-plane spectra is the reduction of the carrier scattering rate \cite{pseudogapinplane}. Therefore, it is possible to observe the pseudogap state more easily in the \textit{c}-axis conductivity ($\sigma_{1,c}(\omega)$) than in the \textit{a}-axis one ($\sigma_{1,a} (\omega)$). Although both the superconducting and the insulating gaps create seemingly similar features, namely the suppression of the low-$\omega$ $\sigma_{1,c}(\omega)$, we can separate these two, based on the behavior of the spectral weight (SW) transfer. For a superconducting gap, the lost SW is redistributed as a $\delta$-function at $\omega=0$, while for an insulating gap, it is transferred to higher energy region.

YBa$_{2}$Cu$_{3}$O$_{7-\delta}$ is the best system to investigate the \textit{c}-axis optical conductivity because it has a relatively high conductivity among all the cuprates. On the other hand, some complicated structures has been observed in this system due to the multilayer structure. In the present work, we study the pseudogap electronic state in a wide energy range, based on a detailed analysis of the SW redistribution with temperature, for the underdoped YBa$_{2}$(Cu$_{1-x}$Zn$_{x}$)$_{3}$O$_{7-\delta}$. The aim of the Zn-substitution is to suppress the so-called transverse Josephson plasma mode \cite{Zn-YBCO} that introduces an additional SW transfer and makes the SW discussion complicated \cite{Zn-YBCO2}. 

YBa$_{2}$(Cu$_{1-x}$Zn$_{x}$)$_{3}$O$_{7-\delta}$ single crystals (for $x=0,$ $0.007$, $0.012$, and $0.04$) were grown by using a pulling technique explained elsewhere \cite{pullingtechnique}. The sample surfaces, which were cut along the \textit{c}-axis (\textit{ac}-plane $\sim2.5$ $\times2.5$ mm$^{2}$), were mechanically polished by using Al$_{2}$O$_{3}$ powder gradually as fine as $0.3$ $\mu$m. In this paper, we focus on the results for the doping levels \textit{p} = $0.11$ and \textit{p} = $0.13$ (estimated from the \textit{p}-\textit{T}$_{c}$ curve \cite{Tcvsp}), where the \textit{T}$_{c}$ values were determined by the dc magnetic susceptibility measurements. The doping level was controlled by annealing under O$_{2}$ flow at $675$ ${{}^\circ}$C and $625$ ${{}^\circ}$C for \textit{p} = $0.11$ and \textit{p} = $0.13$, respectively; followed by a rapid quench into liquid nitrogen. In order to keep the same doping level for all the Zn-content samples, we annealed the samples simultaneously. The \textit{T}$_{c}$ and the pseudogap temperatures \textit{T}$^{\ast}$ are summarized in Table.\ref{samples}. \textit{T}$^{\ast}$ values have been obtained by the SW analysis of our optical data, as explained in detail later, and in agreement with the previously published values \cite{dcresistivityT*}.

\begin{table}[h] \centering
\small\addtolength{\tabcolsep}{-4pt}
\caption{Zn-content (\textit{x}), doping level (\textit{p}), \textit{T}$_{c}$, transition width $\Delta$\textit{T}$_{c}$ and \textit{T}$^{\ast}$}
\begin{tabular}
[c]{ccccc}\hline
{Zn-content (\textit{x})} & {Doping Level (\textit{p})} &
\textit{T}$_{c}$ (K) & $\Delta$\textit{T}$_{c}$ (K) & \textit{T}$^{\ast}$ (K)\\\hline
0 & 0.11 & 61& 3 & 270$\pm20$\\
0.007 & 0.11 & 43 & 7 & 270$\pm20$\\
0.012 & 0.11 & 29 & 7 & 270$\pm20$\\
0 & 0.13 & 71 & 3 & 200$\pm20$\\
0.012 & 0.13 & 37 & 7 & 200$\pm20$\\
0.04 & 0.13 & non-superconducting &  & 200$\pm20$\\\hline
\end{tabular}
\label{samples}
\end{table}

\begin{figure}[ptb]
\centering
\includegraphics[]
{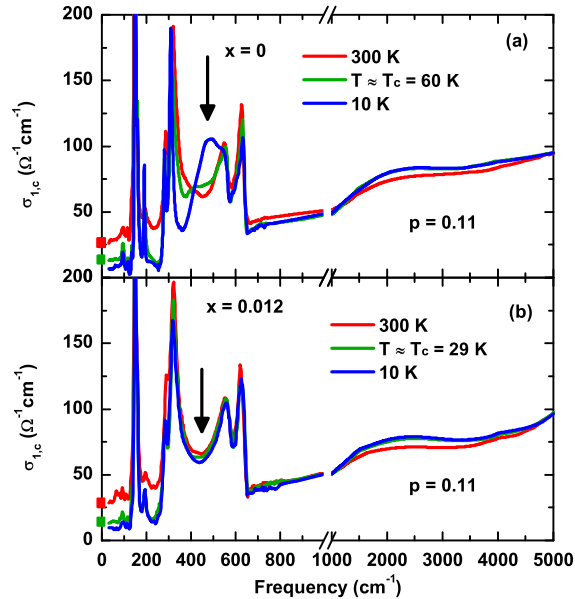}
\caption{(Color online) Temperature dependent high energy optical conductivity  spectra at \textit{T}$^{\ast}$, \textit{T}$_{c}$ and $10$ K for YBa$_{2}$(Cu$_{1-x}$Zn$_{x}$)$_{3}$O$_{7-\delta}$ with (a) $x=0$, (b) $x=0.012$, respectively in  \textit{p} = $0.11$ doping region. Low energy part clearly demonstrate the disappearance of the TJP mode for $x=0.012$.}
\label{comparison}
\end{figure}

The temperature dependent reflectivity spectra were measured with a Bruker 80v Fourier Transform Infrared (FTIR) spectrometer from $\sim20$ to $\sim20000$ cm$^{-1}$ with \textit{E}$\parallel$\textit{c}-axis at various temperatures from $10$ to $300$ K. The sample and the reference mirror were placed into a He-flow cryostat and their spectra were compared successively at each measured temperature by checking their positions with a He-Ne laser. The optical conductivity spectra were obtained from the reflectivity spectra by using the Kramers-Kronig transformation. At the high energy region from $2.5$ eV up to $40$ eV, we used room-temperature reflectivity spectra, which were measured with the use of synchrotron radiation at UV-SOR, Institute for Molecular Science (Okazaki). Above $40$ eV, R $\sim$ $\omega^{-4}$ extrapolation has been adopted. For the low energy extrapolations we fitted our reflectivity data with Drude (for the weak electronic background) and Lorentz oscillators (for the phonon contributions that are dominated the \textit{c}-axis spectra) in the normal state. In the superconducting state, we used two-fluid approximation to fit the electronic response.These extrapolations work well, which can be seen in the coincidence of dc and far-infrared conductivity values. (For example, see Fig \ref{comparison}.). 

\begin{figure}[ptb]
\centering
\includegraphics[]
{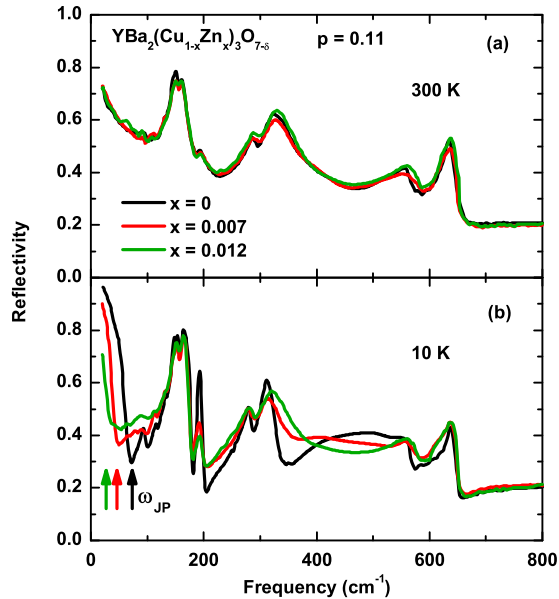}
\caption{(Color online) Comparison of the reflectivity spectra of YBa$_{2}$(Cu$_{1-x}$Zn$_{x}$)$_{3}$O$_{7-\delta}$ at $300$ K (a) and at $10$ K (b) for  \textit{p} = $0.11$. }
\label{spectra}
\end{figure}

Figures \ref{spectra}(a) and (b) show the reflectivity spectra at \textit{p} = $0.11$ for all the Zn-contents at $300$ and $10$ K, respectively. Similar spectral behaviors can be seen also for \textit{p} = $0.13$. At $300$ K, all the spectra are nearly the same. Since the peak intensities of the two phonon modes around $550$ and $625$ cm$^{-1}$ are very sensitive to the oxygen content \cite{PSreliable}, the almost identical spectra for the three samples prove that the doping levels of these samples are the same. 

While there is almost no Zn effect at $300$ K, a clear Zn-effect is observed at lower temperatures. When we cool down the samples below \textit{T}$_{c}$, a sharp Josephson plasma (JP) edge appears below $100$ cm$^{-1}$ (pointed by the arrows in Fig. \ref{spectra}(b)). With increasing Zn-content, the JP edge shifts to the lower energies, indicating the decrease of the superfluid density \cite{Zn-YBCO, Zneffect}. This is naturally expected since the Zn is a well known impurity that causes pair breaking \cite{msr-Zneffect, msr-Zneffect2, optical-Zneffect}. Another significant change is the growth of a broad peak around $\sim450$ cm$^{-1}$. This additional mode has been intensively studied and attributed to the transverse Josephson plasma (TJP) resonance \cite{Zn-YBCO2, TJP resonance}, which appears as a response of the superconducting carriers in the multilayer structure. It is clearly seen that the TJP mode is weakened by Zn-substitution (arrow in Fig. \ref{spectra}(b)), as was reported previously \cite{Zn-YBCO}.

Figures \ref{comparison}(a) and (b) present the optical conductivities up to $\sim5000$ cm$^{-1}$ for $x=0$ and $0.012$, respectively, which were calculated from the reflectivity spectra in Fig \ref{spectra}. The data for the Zn-free sample is in good agreement with the published spectra \cite{PSreliable, PSreliable1, PSreliable2}. Low energy conductivities are close to the dc conductivity values (solid squares), and consistent with the previously reported data \cite{Zn effect}. This guarantees the reliability of our measurements and analyses. Since the TJP mode is completely suppressed by Zn-substitution (Fig. \ref{comparison}(b)), we can more clearly observe the low-$\omega$ conductivity suppression due to the pseudogap for $x=0.012$ than for $x=$ $0$. In the latter ( $x=0$), the grown TJP mode peak seems to compensate for the lost SW at low energies. The  high energy spectra, on the other hand, show a similar behavior for both samples. One can see that the suppressed SW below $\sim1200$ cm$^{-1}$ is transferred to the higher energy region above $\sim1200$ cm$^{-1}$, and that the enhancement of SW extends to a very high energy ($\sim5000$ cm$^{-1}$). 

\begin{figure}[ptb]
\centering
\includegraphics[]
{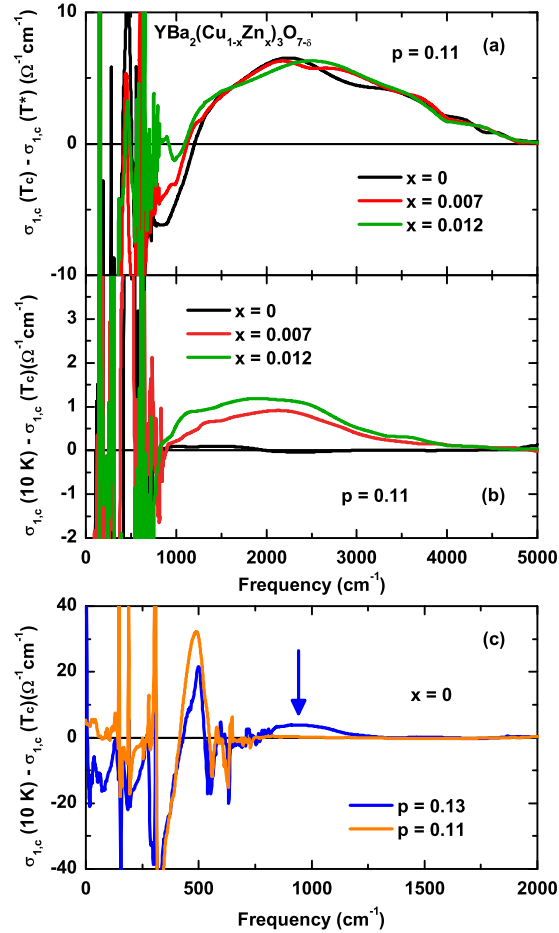}
\caption{(color online) The optical conductivity difference between  \textit{T}$_{c}$ and \textit{T}$^{\ast}$ (a) and between $10$ K and \textit{T}$_{c}$ (b) for $p=0.11$. The bottom figure (c) shows the comparison of the difference spectra between $10$ K and \textit{T}$_{c}$ ($\sigma_{1}(10K)-\sigma_{1}(T_{c})$) at $x=0$ for  $p=0.11$ and  $p=0.13$. Arrow indicate the position of the transverse Josephson plasma frequency at high energy region.}
\label{TJP}
\end{figure}

The SW redistribution is more clearly demonstrated in the difference spectra ($\sigma_{1,c}(T\approx T_{c})-\sigma_{1,c}(T\approx T^{\ast})$) in Fig.  \ref{TJP}(a). One can see that the SW redistribution occurs within the energy range below \ $\sim5000$ cm$^{-1}$ for all the samples. The transferred SW does not change significantly by Zn-substitution. Below \textit{T}$_{c}$ (Fig. \ref{TJP}(b)), on the other hand, the spectrum shows a clear difference for the Zn-free and the Zn-substituted samples. For $\textit{x}=0$, the SW transfer stops at \textit{T}$_{c}$, while it continues even below \textit{T}$_{c}$ for $\textit{x}=0.007$ and $0.012$. The continuous SW transfer below \textit{T}$_{c}$ indicates that the pseudogap survives in the superconducting state. 

At the doping level \textit{p} = $0.11$, the TJP is located near $450$ cm$^{-1}$ giving no effect on the spectra above $800$ cm$^{-1}$. Therefore, we can conclude that the observed SW transfer from low to high energy region can be solely attributed to the pseudogap. On the other hand, this is not the case for \textit{p} = $0.13$. In Fig. \ref{TJP}(c), when $p$ is increased, the TJP mode creates an additional enhancement of $\sigma_{1,c}(\omega)$ around $1000$ cm$^{-1}$ $^{(}$\cite{TJPshift}$^{}$, which complicates the discussion of the pseudogap effect. Therefore, it is important to suppress this mode by Zn-substitution.

\begin{figure}[pb]
\centering
\includegraphics[]
{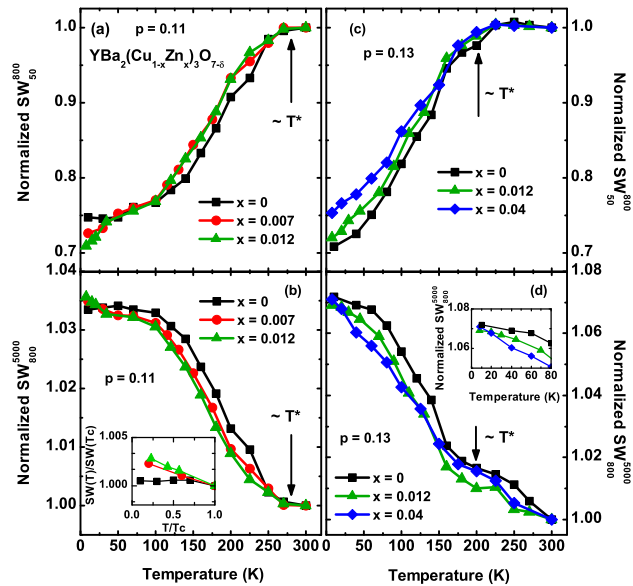}
\caption{(Color online) Temperature dependent SWs normalized at $300$ K for $x=0,0.007,$ and $0.012$ at $p=0.11$. (a) from $50$ to $800$ cm$^{-1}$ (low energy region) and (b) from $800$ to $5000$ cm$^{-1}$ (high energy region). The inset of (b) is the SW$_{800}^{5000}$ normalized at \textit{T}$_{c}$ that demonstrates the SW transfer below \textit{T}$_{c}$ more clearly. (c) and (d) show temperature dependent SWs normalized at $300$ K for $x=0,0.012,$ and $0.04$ at $p=0.13$. (c) from $50$ to $800$ cm$^{-1}$ and (d) from $800$ to $5000$ cm$^{-1}$}
\label{SW}
\end{figure}

To see a more precise temperature dependence, we calculated the \textit{T}-dependent SW normalized at $300$ K for the specific energy regions for \textit{p} = $0.11$ (Fig. \ref{SW}(a) and (b)) and \textit{p} = $0.13$ (Fig. \ref{SW}(c) and (d)). Here, $SW_{a}^{b}=\int_{a}^{b}\sigma_{1,c} (\omega)d\omega$, where `\textit{a}' is the initial value and `\textit{b}' is the final value of wave number for the integral. Hereafter, we discuss the SWs in the two energy regions. The first is the low energy region ($50-800$ cm$^{-1}$), where the SW is suppressed with decreasing $T$. The second is the high energy region ($800-5000$ cm$^{-1}$), where the SW is enhanced with decreasing $T$. The higher limit $5000$ cm$^{-1}$ has been chosen, because all the SW transfer occurs below this energy as can be seen in Fig. \ref{TJP}(a). 

Figure \ref{SW}(a) shows the SW below $800$ cm$^{-1}$, where we can see a gap-like suppression due to both the pseudogap and the superconducting condensation. In this energy region, the suppression starts well above \textit{T}$_{c}$ at the temperature \textit{T}$^{\ast}$ that corresponds to the pseudogap opening. \textit{T}$^{\ast}$ does not change significantly with Zn-substitution as was reported by many groups \cite{changeT*, changeT*1, changeT*2}. Below \textit{T}$_{c}$, a further suppression is observed for the Zn-substituted samples, while the decrease of the SW is small for the Zn-free sample, probably because the developed TJP mode compensates the decrease of the SW due to superconducting condensation. Here we note that the optically determined value of \textit{T}$^{\ast}$ is in good agreement with the dc resistivity results reported previously for this doping level \cite{dc resistivity T*}.

Fig. \ref{SW}(b) plots the SW between the $800$ and $5000$ cm$^{-1}$- high energy range. The temperature dependent behavior is just opposite to that is seen in Fig. \ref{SW}(a), which indicates the decreased SW below $800$ cm$^{-1}$ moves to the SW above $800$ cm$^{-1}$. As pointed out previously \cite{saturationbelowTc}, the fact that SW$_{800}^{5000}$ never decreases below \textit{T}$_{c}$ gives evidence against the precursor scenario that explains the pseudogap as a precursor of superconductivity. In this scenario, the SW that is transferred to the high energy region with the pseudogap opening should come back to a $\delta$-function at $\omega=0$ below \textit{T}$_{c}$. However, such a decrease was not observed for any samples.

In contrast to the Zn-insensitive behavior at high temperatures, the lower-\textit{T} SW shows some Zn-effect such as the continuous increase of SW$_{800}^{5000}$ with decreasing \textit{T}, which can be clearly seen in the inset of Fig. \ref{SW}(b). While for the Zn-free sample, the increase due to the pseudogap saturates below \textit{T}$_{c}$, for the Zn-substituted samples an increase can be seen. This implies that the pseudogap state coexists with the superconducting state below \textit{T}$_{c}$. This coexistence possibly takes place even in the Zn-free sample, if we suppose that the SW saturation below \textit{T}$_{c}$ is due to the strong competition between the decrease due to superconductivity and the increase due to the pseudogap.

The overall spectral behavior is the same for the other doping level   $p=0.013$, as can be seen in Fig. \ref{SW}(c) and (d). One clear difference from $p=0.11$ is that the SW$_{50}^{800}$ starts to decrease at a lower temperature ( Fig. \ref{SW}(c)), as one can expect from the established doping dependence of \textit{T}$^{\ast}$. Another important finding is that the pseudogap behavior (the decrease of SW$_{50}^{800}$ and the increase of SW$_{800}^{5000}$) is observed even in the non-superconducting sample with $x=0.04$ at the same \textit{T}$^{\ast}$ (Fig. \ref{SW}(c) and (d)). This is the strong evidence against the theory that the pseudogap is a precursor of superconductivity.

In Fig. \ref{SW}(d) all the samples show the continuous increase of the high energy SW (SW$_{800}^{5000}$) down to the lowest temperature. For $x=0.012$, this increase can be solely attributed to the pseudogap, and thus implies that the pseudogap remains intact even below \textit{T}$_{c}$. On the other hand, for $x=0$, the increase of SW$_{800}^{5000}$  below \textit{T}$_{c}$ is mainly due to the effect of TJP resonance mode (see Fig. \ref{TJP}(c)). The increasing behavior above \textit{T}$^{\ast}$ seen in Fig. \ref{SW}(d) is not related to the pseudogap, but represents the overall SW gain in the presented energy region, probably due to the weak metallic behavior above \textit{T}$^{\ast}$, which also creates the weak increase in the low energy region (Fig. \ref{SW}(c)) above \textit{T}$^{\ast}$.

The coexistence of the pseudogap and the superconducting gap below \textit{T}$_{c}$ has been suggested by other probes, too. Recent angle resolved photoemission spectroscopy (ARPES) studies on Bi$_{2}$Sr$_{2}$CuO$_{6+x}$ single crystals showed that the portion of the pseudogap remain intact even below \textit{T}$_{c}$ \cite{ARPES-coexist}. Moreover, time-resolved pump-probe optical studies \cite{pump-probe-coexist} and scanning tunneling microscopy/spectroscopy (STM/STS) \cite{STM/STS-coexist} on Bi$_{2}$Sr$_{2}$CaCu$_{2}$O$_{x}$ have revealed the coexistence of the pseudogap and the superconducting gap below \textit{T}$_{c}$. This is attributed to the stabilization of the fluctuating charge order observed as a checkerboard pattern in STM/STS. 

In summary, we studied the temperature dependence of the \textit{c}-axis optical conductivity spectra for YBa$_{2}$(Cu$_{1-x}$Zn$_{x}$)$_{3}$O$_{7-\delta}$ single crystals with various Zn-concentrations and doping levels. It has been revealed that the SW transferred to the higher energy region below \textit{T}$^{\ast}$ never comes back to the $\delta$-function at $\omega=0$ at \textit{T}$_{c}$. This indicates that the pseudogap is not a precursor of superconductivity. This conclusion is also supported by the observation of the pseudogap in the heavily Zn-doped non-superconducting sample. In the Zn-substituted samples, the high energy SW continues to increase even below \textit{T}$_{c}$, which suggests the coexistence of the pseudogap state and the superconducting condensation in the superconducting state. Moreover, the observed SW behavior difference between the Zn-free and the Zn-substituted samples suggests that the pseudogap not only coexists  but also competes with superconductivity. 
\\\\\textbf{Acknowledgements}
\\\
This work was supported by the Grant-in-Aid for Scientific Research from the Ministry of Education, Culture, Sports, Science and Technology of Japan (KIBAN(A) No.19204038). We thank T. Tohyama for his useful discussion.

\bibliography{pseudogap}

\end{document}